\documentclass[11pt]{article}
\usepackage{blois,epsfig}
\usepackage[latin1]{inputenc}
\usepackage[OT1]{fontenc}
\usepackage{graphicx} 

\def\Journal#1#2#3#4{{#1} {\bf #2}, #3 (#4)}

\def\PLB{{\em Phys. Lett.}  B}
\def\PRL{\em Phys. Rev. Lett.}
\def\PRD{{\em Phys. Rev.} D}

\def\be{\begin{equation}}
\def\ee{\end{equation}}
\def\bea{\begin{eqnarray}}
\def\eea{\end{eqnarray}}

\begin{document}
\vspace*{4cm}
\title{The NuMoon experiment: first results}

\author{S. Buitink$^{a}$, J. Bacelar$^b$, R.Braun$^c$, G. de Bruyn$^{d,e}$, H. Falcke$^a$, O. Scholten$^f$, K. Singh$^f$, 
        B. Stappers$^g$, R. Strom$^{d,h}$, R. al Yahyaoui$^{f}$}

\address{\hskip 1cm a. Radboud Univ.~Nijmegen, Department of Astrophysics, IMAPP, P.O.Box 9010,\\ 6500 GL Nijmegen, 
            The Netherlands, e-mail: s.buitink@astro.ru.nl\\
         b. ASML Netherlands BV, P.O.Box 324, 5500 AH Veldhoven, The Netherlands\\
	 c. CSIRO-ATNF, P.O.Box 76, Epping NSW 1710, Australia\\
	 d. ASTRON, 7990 AA Dwingeloo, The Netherlands\\
	 e. Kapteyn Institute, University of Groningen, 9747 AA Groningen, The Netherlands\\
	 f. Kernfysisch Versneller Instituut, University of Groningen, 9747 AA, Groningen, The Netherlands\\
	 g. School of Physics \& Astronomy, Alan Turing Building, Univ. of Manchester, Manchester, M13 9PL\\
	 h. Astronomical Institute `A. Pannekoek', University of Amsterdam, 1098 SJ, The Netherlands\\}

\maketitle
\abstracts{The NuMoon project uses the Westerbork Synthesis Radio Telescope to
search for short radio pulses from the Moon. These pulses are created
when an ultra high energy cosmic ray or neutrino initiates a particle
cascade inside the Moon's regolith. The cascade has a negative charge
excess and moves faster than the local speed of light, which causes
coherent Cherenkov radiation to be emitted. With 100 hours of data, a
limit on the neutrino flux can be set that is an order of magnitude
better than the current one (based on FORTE). We present an analysis
of the first 10 hours of data.
}

\section{Introduction}
Above the Greisen-Zatsepin-Kuzmin (GZK) energy of $6\cdot 10^{19}$~eV, cosmic rays (CRs) can interact with the cosmic 
microwave background photons, losing energy and producing pions when traversing distances of the order of 10
Mpc \cite{g66,zk66}. 
Recent results of the Pierre Auger Observatory have confirmed a steepening in the cosmic ray spectrum at the GZK energy \cite{Y07}. 
The pions will produce ultrahigh energy (UHE) neutrinos through weak decay.
Observing these neutrinos is of great scientific interest as their arrival direction
points back to sources at distances larger than 10 Mpc, 
in contrast to charged particles that are deflected in (extra-)galactic
magnetic fields. In addition, while cosmic rays from faraway sources pile up at the GZK energy, information
about the cosmic ray spectrum at the source is conserved in the GZK neutrino flux. 
Other possible sources of UHE neutrinos are decaying supermassive particles, such as magnetic monopoles or 
topological defects. This class of models is refered to as top-down (TD) models (see for example Stanev \cite{s04} for 
a review). 

Because of their small interaction cross section and low flux, the detection of cosmic neutrinos 
calls for extremely large detectors. Assuming the Waxman-Bahcall flux \cite{wb01}, even at GeV energies the flux is not
higher than a few tens of neutrinos per km$^2$ per year. Kilometer-scale detectors are not
easily built but can be found in nature. For example, interaction of neutrinos in ice or water
can be detected by the Cherenkov light produced by the lepton track or cascade. The now half finished
IceCube detector \cite{icecube} will cover a km$^3$ volume of South Pole ice with PMTs, while Antares \cite{antares} 
and its successor KM3NET \cite{KM3NET}
exploit the same technique in the Mediterranean sea. Even larger volumes can be covered by
observing large detector masses from a distance. The ANITA balloon mission \cite{anita} monitors an area of
a million km$^2$ of South Pole ice from an altitude of $\sim 37$~km and the FORTE satellite \cite{forte} 
can pick up radio signals coming from the Greenland ice mass. Alternatively, cosmic ray experiments like the Pierre Auger
Observatory can possibly distinguish cosmic ray induced air showers from neutrino induced cascades at very
high declinations where the atmosphere is thickest and only neutrinos can interact close to the detector.

For an even larger detector volume one can turn to the Moon. The negative charge excess of a particle shower inside 
a dense medium will cause the emission of coherent Cherenkov radiation in a process known as the Askaryan
effect \cite{a62}. This emission mechanism has been experimentally verified at accelerators \cite{s01,g00} and extensive
calculations have been
performed to quantify the effect \cite{zhs92,az97}. The idea to observe this type of emission from the Moon
with radio telescopes was first proposed by Dagkesamanskii and Zheleznyk \cite{dz89} and the first experimental endeavours in this 
direction were carried out with the Parkes telescope \cite{parkes} and at Goldstone (GLUE) \cite{glue}.

The NuMoon project uses the Westerbork Synthesis Radio Telescope (WSRT) to watch for the same flashes but at lower frequency, which has the distinct
advantage that radio pulses have a much higher chance of reaching the observer,
which will be explained in the next section.
  
\section{Theory}
UHE neutrinos or CRs interact below the lunar surface. About 20\% of the neutrino energy is converted into a hadronic shower,
while the other 80\% is carried off by the lepton, which will not produce any observable radio emission. Muons will not produce
enough charge density, while electromagnetic showers become elongated at high energies due to the
Landau-Pomeranchuk-Migdal effect \cite{lpm}. For these showers the angular spread of the radio emission around the Cherenkov
angle becomes very small, severely lowering the chance of detection.

\begin{figure}
%\rule{5cm}{0.2mm}\hfill\rule{5cm}{0.2mm}
%\vskip 2.5cm
%\rule{5cm}{0.2mm}\hfill\rule{5cm}{0.2mm}
\psfig{figure=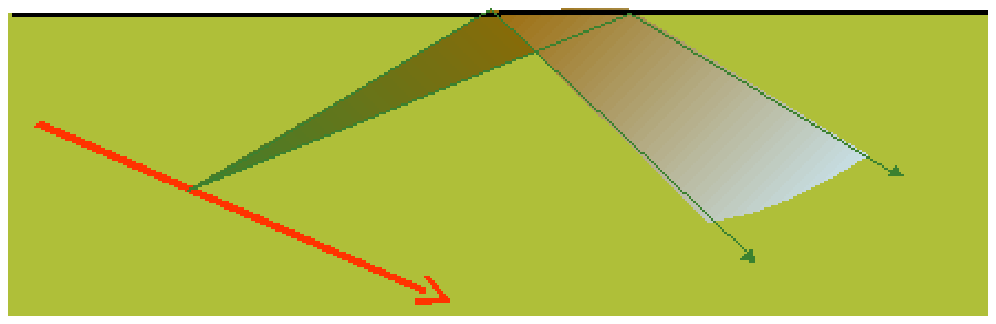, width=.45\linewidth}
\hskip 1cm
\psfig{figure=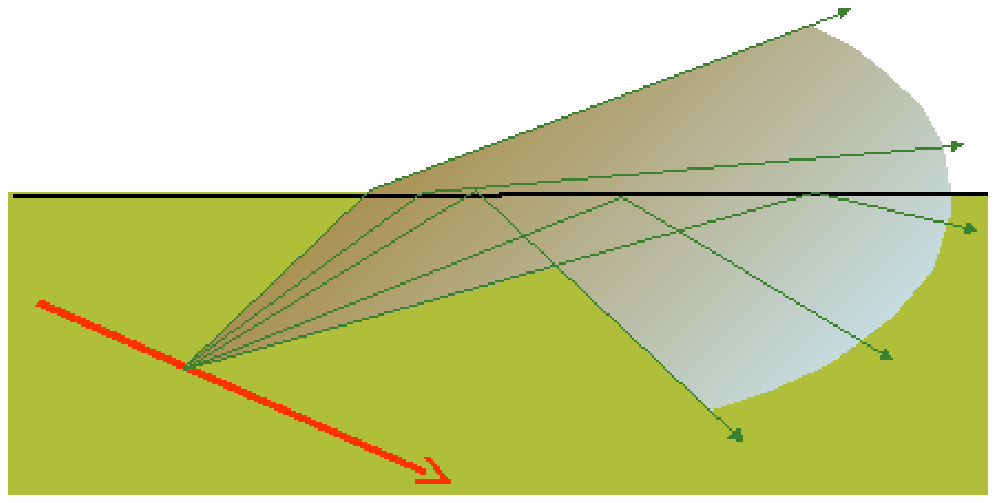, width=.45\linewidth}
\caption{Propagation of radio waves inside the Moon. Left: at 2 GHz the radiation is emitted only close
to the Cherenkov angle and is unable to escape the Moon (for this geometry). Right: at lower frequencies the
angular spread of the radiation is much larger and the chance for radio waves to reach the observer is
increased.}
\label{fig:geometry}
\end{figure}

The lateral size of the cascade is of the order of 10~cm so the radio emission is coherent up to $\sim 3$GHz. Former
experiments, like GLUE, have observed at high frequencies (2.2~GHz) where the emission is strongest. For lower frequencies, however, the
angular spread of the emission around the Cherenkov angle increases. For
emission at the Cherenkov angle only showers hitting the rim of the Moon under such an angle that the emission will not be
internally reflected at the Moon surface will be observable. With a larger angular spread in the emission a wider range of
geometries is allowed (see Fig. \ref{fig:geometry}) and a larger part of the lunar surface can be used. When the wavelength is of the order of the shower
length, several meters, the emission becomes nearly isotropic and pulses can be expected to come from the whole Moon \cite{scholten}. 
In our experiment we exploit this optimal frequency range around 150 MHz. 

The intensity of the radio emission from a hadronic shower with energy $E_s$ in the lunar regolith can be parameterized
as \cite{zhs92,az97,scholten}:
\begin{eqnarray}
F(\theta,\nu,E_s)=3.86\cdot 10^4 \exp^{-Z^2} \left(\frac{\sin\theta}{\sin\theta_c}\right)^2
\left(\frac{E_s}{10^{20}\mathrm{eV}}\right)^2
\left(\frac{d_{\mathrm{moon}}}{d}\right)^2 \nonumber \\
\left(\frac{\nu}{\nu_0 (1+(\nu/\nu_0)^{1.44})}\right)^2
\left(\frac{\Delta \nu}{\mathrm{100 MHz}}\right) \mathrm{Jy}
\end{eqnarray}
with:
\begin{equation}
Z=(\cos\theta-1/n)\left(\frac{n}{\sqrt{n^2-1}}\right)\left(\frac{180}{\pi\Delta_c}\right)
\end{equation}
where $\Delta \nu$ is the bandwidth, $\nu$ the central frequency and $\nu_0=2.5$~GHz. The average Earth-Moon distance
$d_\mathrm{moon}=3.844\cdot 10^{8}$~m, $d$ is the distance to the observer. The Cherenkov angle is given by
$\cos\theta_c=1/n$, where $n$ is the index of refraction and $\theta$ is the angle under which radiation is emitted
relative to the direction of shower propagation. The spread of radiation around the Cherenkov angle is given by:
\begin{equation}
\Delta_c=4.32^{\circ}\left(\frac{1}{\nu[\mathrm{GHz}]}\right)\left(\frac{L(10^{20}\mathrm{eV})}{L(E_s)}\right)
\end{equation}
where $L$ is the shower length depending on primary energy.
 
The regolith is the top layer of the Moon and consists of dust and small rocks. The properties of this layer are known from
samples brought from the Moon by the Apollo missions \cite{os75}. The average index of refraction is $n=1.8$ and the
attenuation length is found to be $\lambda_{r}=(9/\nu[\mathrm{GHz}])$~m for radio waves \cite{os75,hvf91}. The thickness of the
regolith is known to vary over the lunar surface. At some depth there is a (probably smooth)
transition to solid rock, for which the density is about twice that of the regolith. In Scholten et al.
\cite{scholten} the effects of pure rock and regolith are simulated and found to give very similar detection
limits for low frequencies.

As the radiation leaves the Moon it refracts on the surface, so surface irregularities may play a role
in the detectability. James et al. \cite{james} argue that small scale irregularities may increase detectability
while large scale structures will reduce the chance on detection of CR showers, because they develop
directly underneath the surface and tend to have geometries less favourable for detection. These effects become
less important for low frequencies and large angular spread of the emission.

\section{Experimental setup}
The Westerbork Radio Synthesis Telecope (WSRT) is an array telescope consisting of 14 parabolic
telescopes of 25~m on a 2.7~km east-west line. The NuMoon experiment uses the Low Frequency Front Ends (LFFEs)
which cover the frequency range 115--180 MHz. Each LFFE records full polarization data.
For our observations we use the Pulsar Machine II
backend \cite{kss08}, which can record a maximum bandwidth of 160 MHz sampled as 8 subbands of 20 MHz each. We use
two beams of 4 bands each, centered around 123, 137, 151, and 165 MHz. The two beams are aimed at different 
sides of the Moon, both covering about one third of the lunar surface, in order to
enlarge the effective apperture and create the possibility of an anti-coincidence trigger. A real
lunar Cherenkov pulse should only be visible in one of the two beams. Because of overlap in the subbands the
total bandwidth per beam is 65 MHz. 

For each subband, the time series data is recorded at several storage nodes with a sampling frequency
of 40 MHz.    
\section{Data analysis}
\begin{figure}
%\rule{5cm}{0.2mm}\hfill\rule{5cm}{0.2mm}
%\vskip 2.5cm
%\rule{5cm}{0.2mm}\hfill\rule{5cm}{0.2mm}
\psfig{figure=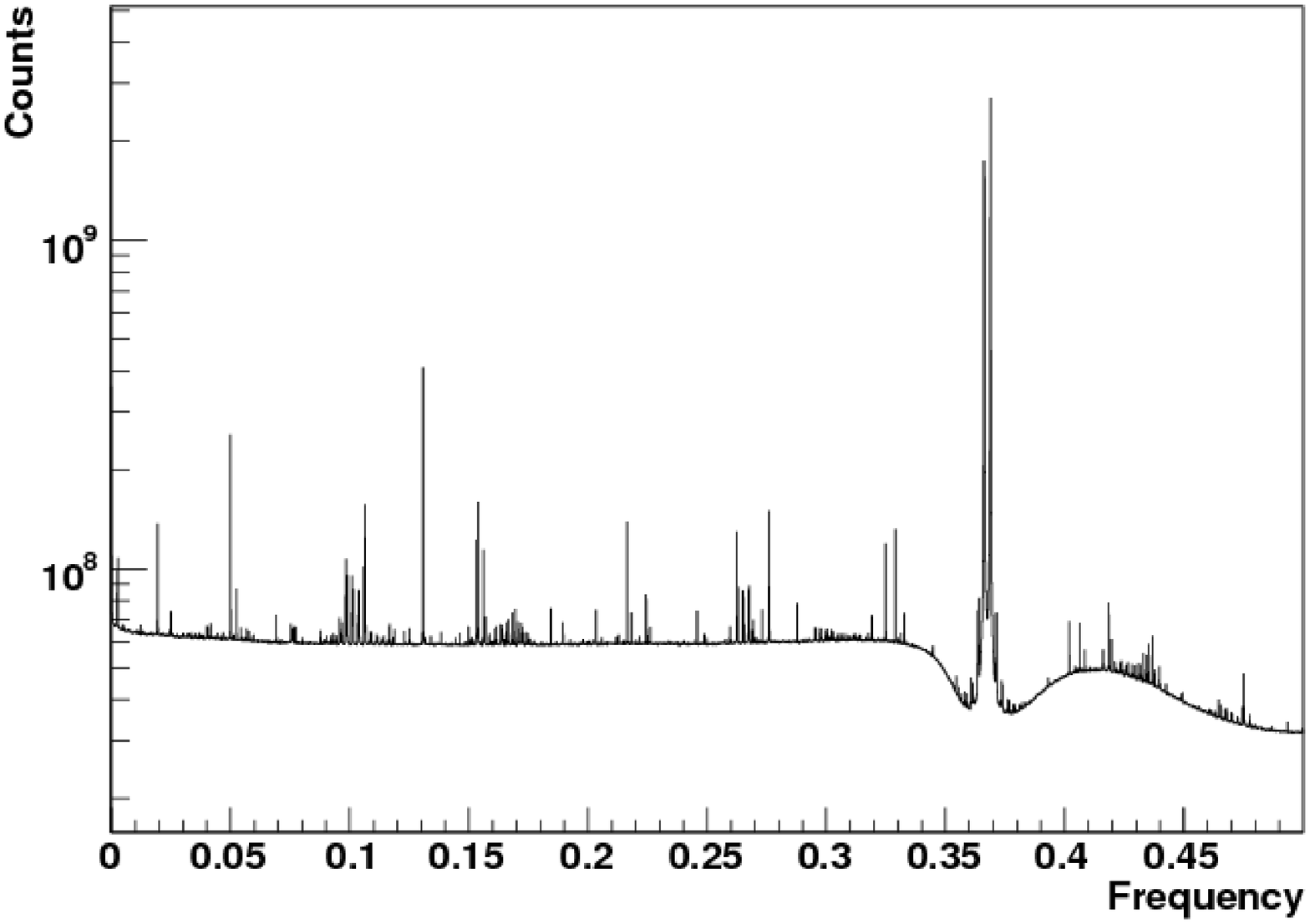, width=.4\linewidth, angle=270}
%\hskip 1cm
\psfig{figure=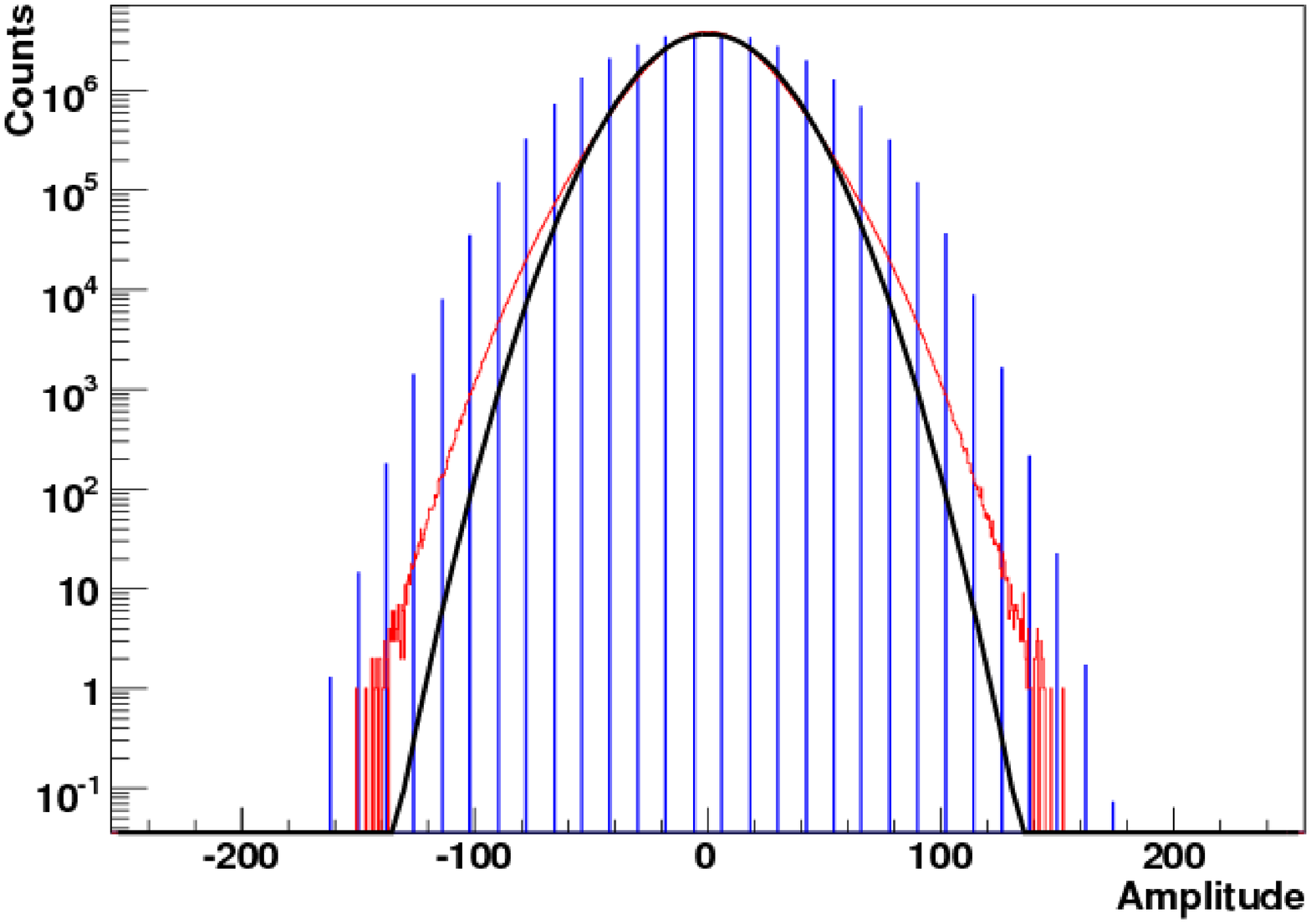, width=.4\linewidth, angle=270}
\caption{Left: Typical frequency spectrum of 10 seconds of data before RFI reduction. The x-axis shows arbitrary units,
corresponding to a bandwidth of 20 MHZ centered around 165 MHz. Right: Typical results of RFI
reduction. The number of counts per amplitude is plotted for the raw data (blue) and the data after RFI reduction (red).
Notice that the raw data has a limited dynamic range. The black line is a Gaussian fit to the data after RFI reduction.}
\label{fig:analysis}
\end{figure}

In the first phase of the data reduction a search for pulses is done to select about 1\% of the data,
which will be permanently stored. The procedure involves the following steps:
\begin{itemize}
\item  The raw data is read in blocks of 20\,000 timebins. A Fourier transform of these blocks is made
to produce a frequency spectrum in which narrow-band radio frequency interference (RFI) can be identified. For this
background reduction 200 spectra are averaged and fitted. RFI lines with a value above 1.5 times the fit
to the average spectrum are suppressed. The left panel in Fig. \ref{fig:analysis} shows a typical frequency spectrum of
10 seconds of data before RFI removal. This particular spectrum is for the band centered around 165 MHz. The right panel
shows the amplitude distribution of the data before (blue) and after (red) RFI removal. Notice that the raw data has a
limited dynamic range. The black line is a Gaussian fit to the data after RFI removal.
\item Pulses from the Moon are dispersed in the Earth's ionosphere. A dedispersion is performed, based on
the total electron content (TEC) of the ionosphere and afterwards the data is transformed back to the time domain.
\item We expect our pulses to have a width smaller than the binsize (25~ns). Generally, such a pulse
will spread over a few timebins, depending on its phase. Errors in the TEC value used for dedispersion will further broaden
the pulse. We define the value $P5$ as the power integrated over 5 timebins and 2 polarizations, normalized to the average
value of this integration:
\begin{equation}
P5=\frac{\displaystyle\sum_{\mathrm{5\ bins}} P_x + P_y}{\bigg<\displaystyle\sum_{\mathrm{5\ bins}} P_x + P_y\bigg>}
\end{equation}
where the averaging is done over 20\,000 timebins. 
\item A peak search is carried out, where we use a trigger condition of a $P5$ value of 2.5 in all 
four frequency bands. A time difference between the peaks in the different
frequency bands is allowed (and expected) due to ionospheric dispersion.
\item For each trigger the data blocks in which the pulses are found are stored for postprocessing.
\end{itemize}
This analysis is done for the two beams separately, but when a pulse is found in one of the beams,
the raw data is stored for both beams and all subbands. In order to reproduce the RFI line suppression the values 
of the suppressed frequencies are also stored.

In the second phase of processing further cuts are applied to the data. Pulses of a width exceeding
8 bins are rejected, as well as pulses that are part of a series of consequent pulses.
  
\section{Results}
\begin{figure}
%\rule{5cm}{0.2mm}\hfill\rule{5cm}{0.2mm}
%\vskip 2.5cm
%\rule{5cm}{0.2mm}\hfill\rule{5cm}{0.2mm}
\centerline{\psfig{figure=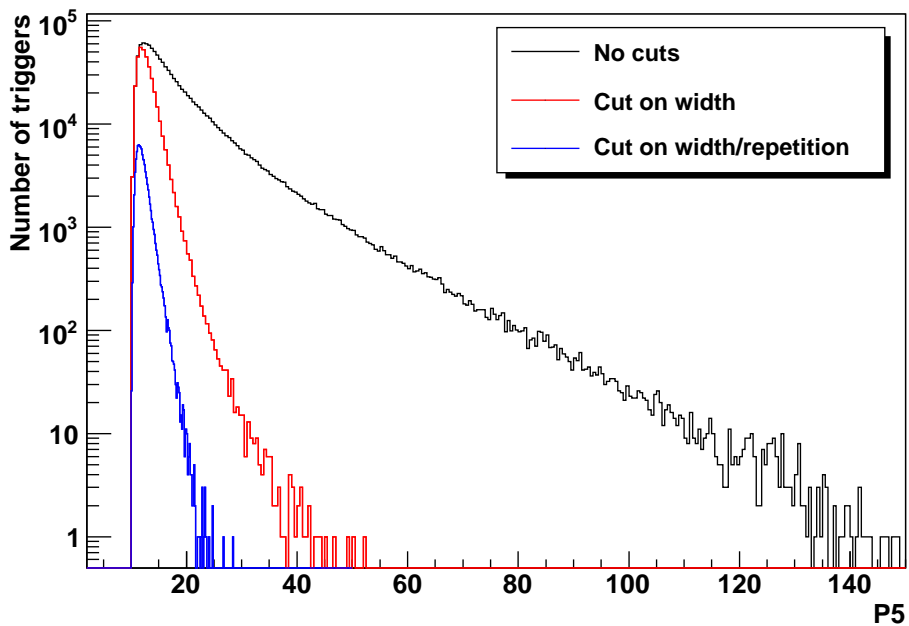, width=.85\linewidth}}
\caption{Number of triggers plotted against the P5 value for different cuts.
\label{fig:cuts}}
\end{figure}
Presently, 10 hours and 40 minutes of single beam data have been accumulated.
Fig. \ref{fig:cuts} shows the distribution of $P5$ values (summed over the 4 subbands) with no cuts applied (black line), a cut on pulse 
width (red line) and a cut that in addition suppressed pulses that are part of a pulse train (green
line). 
For pure Gaussian noise the number of expected triggers is around $10^5$ in total and of the order of unity at
$P5\approx 14$.
The number of triggers after a cut on width and repetition is 87000.

\begin{figure}
%\rule{5cm}{0.2mm}\hfill\rule{5cm}{0.2mm}
%\vskip 2.5cm
%\rule{5cm}{0.2mm}\hfill\rule{5cm}{0.2mm} 
%\includegraphics[width=.5\linewidth]{limits.pdf}
\centerline{\psfig{figure=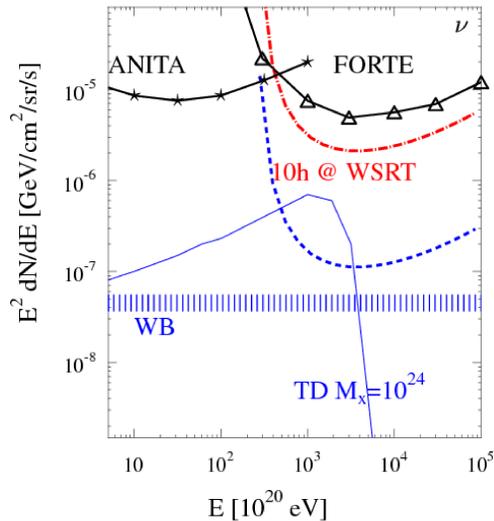, width=.5\linewidth}}
\caption{Neutrino flux limit currently established with 10 hours of WSRT data and the limit that will be achieved
with 100 hours. Limits from ANITA and FORTE 
are included in the plot as well as the Waxman-Bahcall flux and a TD model prediction.
\label{fig:limit}}
\end{figure}

The highest value surviving this cut is $P5=25$ or $P5=6.25$ per frequency subband. For the WSRT the system
noise at low frequencies is $F_{noise}=600$~Jy per polarization channel, so $P5=1$ corresponds to 6000~Jy, giving a 
detection treshold of $\sim 38$~kJy.
Fig. \ref{fig:limit} shows the limit on the neutrino flux that is obtained with the current 10 hours of data based on this detection treshold, a
central frequency of 140~MHz, a bandwidth of 65 MHz, and a coverage of one third of the Moon (corresponding to one beam). Also
shown is the limit that can be achieved with a 100 hour observation period. The current limits in the UHE region
are established by ANITA \cite{anita} and FORTE \cite{forte}. 
In the bottom of Fig. \ref{fig:limit} two model
predictions are plotted: the Waxman-Bahcall limit \cite{wb01} and a top-down
model \cite{ps96} for exotic particles of mass $M_X=10^{24}$~eV.      
\section{Discussion}
We observe at a frequency window that offers an optimal sensitivity to lunar pulses. At the same time, for these low
frequencies the effects of uncertainties concerning the lunar surface and interior are small. Because
of the large spread in emission angle, we expect no systematic effect from surface irregularities. The
detection efficiency is also largely independent from details in the structure of the (sub-)regolith
\cite{scholten}.

A possible complication is the limited dynamic range of the WSRT data. The individual dishes give a
2-bit signal, so the total dynamic range is $3N+1$ for $N$ dishes. When there is a lot of RFI background,
as much as 90\% of the total received power can be in narrow RFI lines. After background reduction the
dynamic range is suppressed and detection of pulses may become impossible. Another effect of the limited
dynamic range is that large pulses will be cut off to the maximum allowed value, potentially lowering
the signal-to-noise. Here, the atmospheric dispersion contributes positively by spreading the power over
more timebins.

Currently we are simulating the detection efficiency for various radio backgrounds and atmospheric
conditions. The error on the TEC value, used for dedispersion, is also taken into account.

\section{Outlook}
With about 100 hours of data and the results of aforementioned simulations, we will be able to put a limit on the UHE
neutrino flux which is about an order of magnitude lower than the current FORTE limit (assuming no detections). This
limit would rule out a subset of TD models (see Fig. \ref{fig:limit}). 
\begin{figure}
%\rule{5cm}{0.2mm}\hfill\rule{5cm}{0.2mm}
%\vskip 2.5cm
%\rule{5cm}{0.2mm}\hfill\rule{5cm}{0.2mm}
\psfig{figure=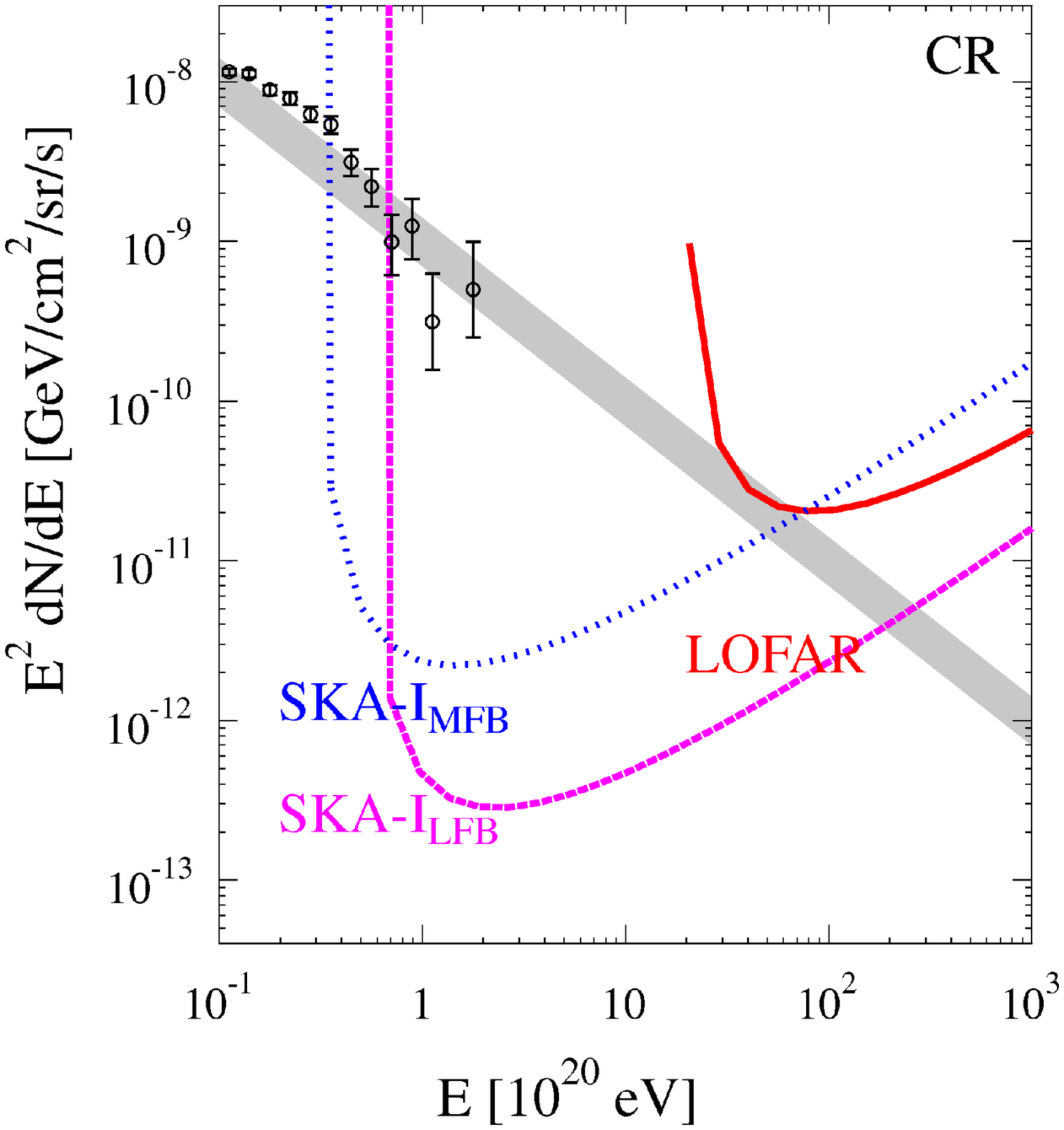, width=.45\linewidth}
\psfig{figure=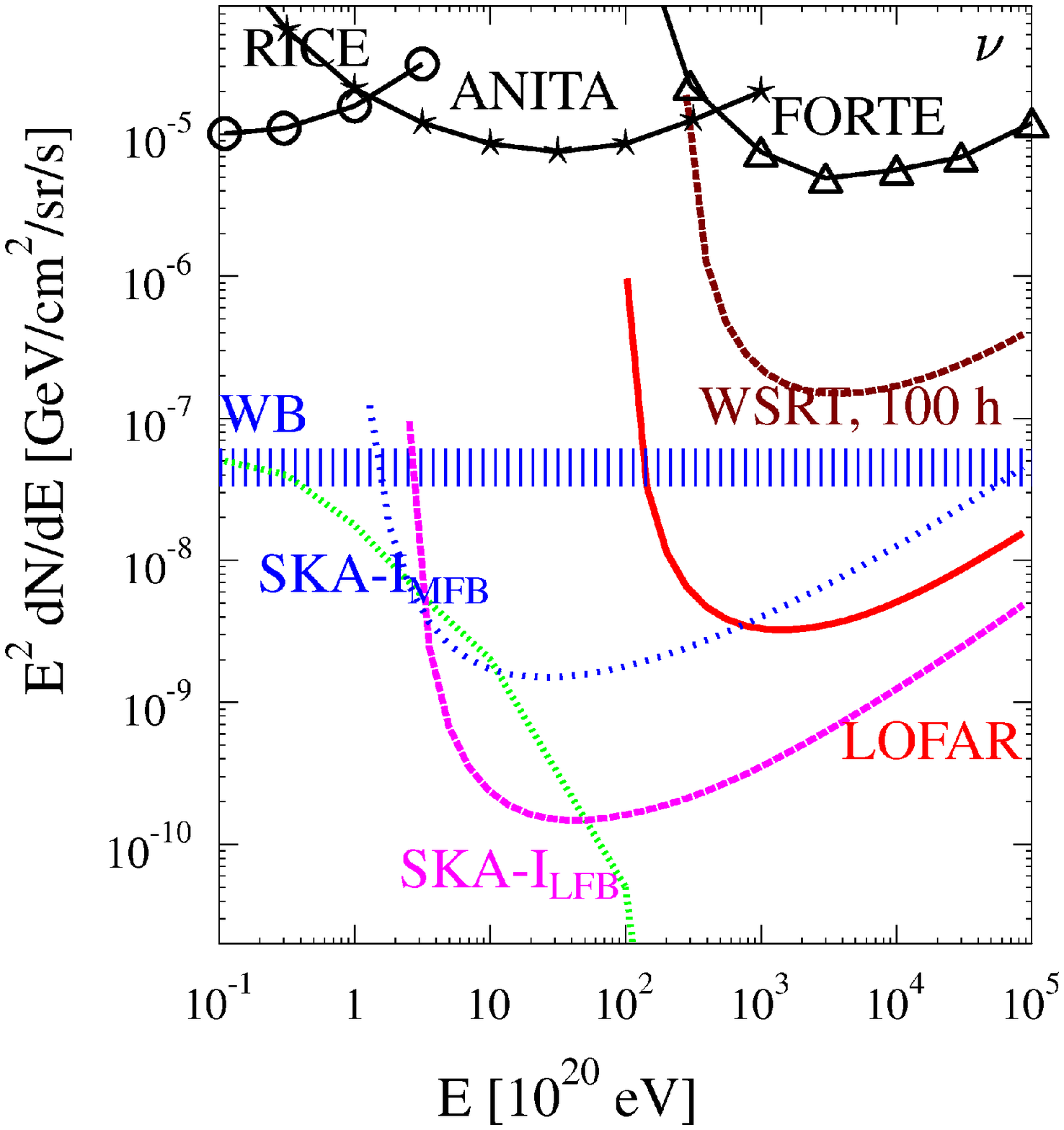, width=.45\linewidth}
\caption{Limits on UHE neutrino flux (left) and cosmic ray flux (right) that can be established with LOFAR and
SKA.}
\label{fig:lofar}
\end{figure}

The next phase in the NuMoon experiment will be to use Lofar, the Low Frequency Array, that is under construction in 
the Netherlands. Lofar is a network of low frequency omni-directional radio antennas communicating over a fiber optics
network. It will feature two types of antennas operating at different frequencies:
the Low Band (LB) antennas cover a band of 30--80~MHz while the High Band (HB) antennas cover
the regime 110--240~MHz. The latter will be used for the NuMoon observations.
Lofar is organized in 35 stations each containing 48 LB and 96 HB antennas. Half of the
stations are located inside the 2~km$\times$2~km core with a total collecting area of $\sim$0.05~km$^2$.
Multiple beams can be formed to cover the surface of the Moon, resulting in a
sensitivity that is about 25 times better than the WSRT.  

The Lofar stations communicate via a central processor (CEP). At the CEP a
real-time trigger system can be implemented combining the incoming data from the
different stations. Locally, the data stream of all antennas is buffered on
Transient Buffer Boards (TBB) for $\sim 1$~s. After a trigger has been found 
the system can read out the TTBs and store the complete time series data of all 
antennas for off-line analysis.   

Fig. \ref{fig:lofar} shows the sensitivity that will be achieved with 30 days of
observation time with Lofar for UHE cosmic rays (left panel) and neutrinos (right panel) \cite{design}.

Other lunar Cherenkov observations will be carried out at the Australia Telescope Compact Array (ATCA) \cite{james} consisting of six 22~m
dishes. The array is currently undergoing an upgrade to be able to measure with a bandwidth of 2 GHz. The upgrade is
projected to be finished in 2009. 

Eventually, the best sensitivity will be achieved with the Square Kilometer Array  \cite{ska}
(SKA), planned to be completed in 2020. The Australian SKA Pathfinder (ASKAP) is expected to be operational around 2011.
A study of the sensitivity of ATCA, ASKAP and SKA can be found in James et al. \cite{james} (note that the LOFAR sensitivity in
their plots is exagerrated). In Fig. \ref{fig:lofar} the expected sensitivity of SKA is plotted for observations 
in the low frequency band (70--200~MHz) and the middle frequency band (200--300~MHz).

\end{document}